\documentclass{epl2}
\usepackage{amssymb}

\title{Scale-Freeness for Networks as a Degenerate Ground State: A Hamiltonian Formulation}
\shorttitle{Scale-Freeness for Networks $\cdots$}
\author{P. Minnhagen\inst{1,2}, S. Bernhardsson\inst{1,2}
\and B. J. Kim\inst{3}}
\shortauthor{P. Minnhagen {\it et. al.}}
\institute{
\inst{1} Department of Physics, Ume{\aa} University, 90187 Ume{\aa}, Sweden  \\
\inst{2} Center for Models for Life, 2100 Copenhagen, Denmark \\
\inst{3} Department of Physics, BK21 Physics Research Division,  and Institute of Basic Science, Sungkyunkwan University, Suwon 440-746, Korea}


\abstract{
The origin of scale-free degree distributions in the context of networks is
addressed through an analogous non-network model in which the node degree
corresponds to the number of balls in a box and the rewiring of links to balls
moving between the boxes.
A statistical mechanical formulation is presented
and the corresponding Hamiltonian is derived.
The energy, the entropy, as well as the degree distribution and 
its fluctuations are investigated at various temperatures. 
The scale-free distribution is shown to correspond to the 
degenerate ground state, which has small fluctuations in the
degree distribution and yet a large entropy. We suggest
an implication of our results from the viewpoint of the stability
in evolution of networks.
}

\begin{document}

\maketitle

Complex networks have undergone a rapid surge of interest and a number of
review articles already exist~\cite{albert02,reviews}. A striking observation in this
field is that many real networks have a broad scale-free like
degree distribution. Why is this and what does it imply for the evolution
mechanism of the networks? The preferential attachment~\cite{barabasi99}
has been proposed as such a mechanism, where the scale-freeness is directly
linked to the growing number of nodes in the network: once a link is in place
it stays in place. This model mechanism has become a successful prototype
explanation in many cases~\cite{albert02,reviews}. In the present context, it
corresponds to the extreme limit with only growing and no rewiring of links~\cite{footnote},
whereas the opposite extreme is no growing and only rewiring. The fact that
also the latter extreme can lead to scale-free degree distributions was
explicitly demonstrated by the discovery of the merging-type 
evolution~\cite{beom05,thurner05}, or the non-growth network
model~\cite{jensen}. Furthermore, one can devise a scale-free
evolution as an in-between of these two extremes. For example,
you can combine
slow growing with the merging type evolution and still get scale-free
distributions~\cite{beom05} and you can combine preferential growing with
deletion of nodes and also get broad degree distributions for somewhat less
fast growing networks~\cite{gronlund04}. However, the distinction remains: In
the one case the scale-freeness is linked to the network growing process and
in the other to the rewiring of links.

In the present Letter, we try to get a better understanding of
scale-freeness through rewiring by using a statistical mechanical
formulation, applied for a non-network model without growing.
We characterize this model in terms of a ground state energy, 
a temperature, and an entropy. \textit{It is found that the scale-free distribution 
is unique in the sense that it combines small deviations from 
the average degree distribution with a high entropy, or equivalently 
a large number of available states.}

Our model consists of $N$ ``towns'' (=boxes) with the total of $N\langle
k\rangle $ ``inhabitants'' (=balls) 
with  $\langle k \rangle$ being the average number of inhabitants
per town~\cite{Burda}. 
We assume that a randomly picked person $A$ chooses a person $B$ 
randomly and then moves to the $B$'s town with a certain probability.
The rate of moving from a town with $k_{1}$ inhabitants to one 
with $k_{2}$ inhabitants, $R(k_{1},k_{2})$, is assumed to be
separable, i.e.,  $R(k_{1},k_{2})=g(k_{1})f(k_{2})$, which means 
that the attraction or repulsion of a town of a certain size is the same
for all people. The question is then what distribution of town sizes this
leads to in a steady state. 
There is a one-to-one correspondence between this model and a corresponding
network: The inhabitants corresponds to link-ends so that each link is
associated with two specific persons. The link-ends are randomly chosen and
randomly moved to other link-ends and attached to the same node. However, in
networks one usually also introduces the topological constraints  a) the
network is connected, b) only one link connects two different nodes, and c) no
link starts and ends on the same node. In practice these topological
constraints only introduces significant statistical differences between the
network and the town model for the case when the random rewiring in steady
state leads to nodes with of order $N\langle k\rangle$ inhabitants\cite{chaos}.

The probability $P(k,t)$ that
a town has $k$ inhabitants at time $t$ evolves following the master equation 
\begin{eqnarray}
N\frac{\partial P(k,t)}{\partial t} =&  -{\cal F}(k,t)f(k)+ {\cal F}(k-1,t)f(k-1) \nonumber \\
&-{\cal F}(k,t)g(k) + {\cal F}(k+1,t)g(k+1), \label{detailed}
\end{eqnarray}
where ${\cal F}(k,t) \equiv kP(k,t)/[\sum_k k P(k,t)] =
kP(k,t)/\langle k \rangle$ is the probability of choosing a person from a
town of size $k$.
The two terms with the negative signs in Eq.~(\ref{detailed}) 
describe the decrease of $P(k)$ when a person moves in and out of a
town of size $k$, respectively. Note that this for a network corresponds to rewiring one link at a time~\cite{footnote2}.
The two functions $f(k)$ and $g(k)$
are required to satisfy the condition 
$\sum_{k}kP(k)f(k)/\langle k \rangle= \sum_{k}kP(k)g(k)/\langle k \rangle=1$ 
in order to ensure that precisely one
person on the average moves at each time step. 
The second and the last terms correspond to when a person moves from 
and to a town of size $k+1$ and $k-1$, respectively, which increase
the inhabitants in the towns of size $k$. 
From the steady state condition together with the normalization
conditions one obtains the detailed balance conditions for $f(k)$ 
for moving into a town, and $g(k)$ for moving out of a town. 
There is in fact only one possibility:
\begin{eqnarray}
f(k)&=&\frac{(k+1)P(k+1)}{\left[1-\frac{P(1)}{\langle k\rangle}\right]kP(k)}\mbox{ \ \ for $k \geq 1$}, \label{f(k)}\\
g(k)&=&\frac{1}{1-\frac{P(1)}{\langle k \rangle}}\mbox{ \ \ for}\ k\geq2, \label{eq:gk}
\end{eqnarray}
and $f(0)=0$ and $g(1)=0$.
In this model, a sole inhabitant is not allowed to leave the town [$g(1) = 0$],
and you never move to an empty town because there is none to encounter [$f(0) =
0$]. We stress that Eqs.~(\ref{detailed})-(\ref{eq:gk}) constitute the only possibility under the given
assumptions.

The next step is to choose an update rule which is consistent with the above
conditions: We start from two random persons $A$ and $B$. Suppose that $A$
($B$) lives in a town of size $k_1$ ($k_2$). We
then also randomly choose two persons $C$ and $D$ who live 
in towns of size $k_{2}+1$ and $k_{1}-1$, respectively. A possible update rule
is that $A$ $either$ moves to $B$'s town $or$ $C$ moves to $D$'s
town. We know from the detailed balance that the number of $A$'s 
moving to $B$'s is on the average equal to the number of $C$'s 
moving to $D$'s. The probability
for $A$ to move to $B$'s town using this update is given by $\frac{f(k_{2}%
)}{f(k_{2})+f(k_{1}-1)}$ and the probability for $C$ moving to $D$'s town is
then $\frac{f(k_{1}-1)}{f(k_{2})+f(k_{1}-1)}$. The latter probability is then
within this update rule equivalent to $A$ $not$ moving to $B$. Next, one
notes that if we refer these two probabilities to moving or not moving $A$ to
$B$'s town, respectively, then detailed balance is in fact automatically
fulfilled. So our particular update rule is equivalent to randomly choosing
two persons $A$ and $B$ and then moving $A$ to $B$'s town with the probability
$\frac{f(k_{2})}{f(k_{2})+f(k_{1}-1)}$. The $relative$ probability for
$A$ to move or not to move to $B$'s town, respectively, can consequently be
expressed in terms of the Boltzmann-type factor $e^{\ln f(k_{2})-\ln
f(k_{1}-1)}$. We rewrite this as $e^{-\Delta E}$ with $\Delta E_{1\rightarrow
2}=\ln f(k_{1}-1)-\ln f(k_{2})=\ln k_{1}P(k_{1})-\ln(k_{1}-1)P(k_{1}%
-1)-[\ln(k_{2}+1)P(k_{2}+1)-\ln k_{2}P(k_{2})]$ which is equivalent to
assigning the energy $\epsilon(k)=\sum_{i=1}^{k}[\ln(iP(i))-\ln
((i-1)P(i-1))]=\ln kP(k)$ to a town of size $k$. This means that the total
energy of the Town model is 
\begin{equation}
E=-N\sum_{k=1}P(k)\ln [kP(k)],
\end{equation}
where $NP(k)$ is the average number of towns with $k$ inhabitants. We note that
this  implies that any predetermined distribution $P(k)$ can be recovered
from a Monte Carlo (MC) algorithm with the canonical distribution 
$e^{-\hat{E}/T}$ with
$T=1$ and $\hat{E} = -N \sum_{k=1} n(k) \ln[ kP(k)]$,
where $Nn(k)$ is the number of towns with $k$ 
inhabitants~\cite{burda01,berg02}. The point in the present context is that
$\hat{E}$ is the Hamiltonian for a given fixed $P(k)$ which yields
the correct average energy $\langle \hat{E} \rangle = 
E = -N\sum_{k=1}P(k)\ln [kP(k)]$.

We also note that the relation between the entropy $S$ and the
probability distribution $P(k)$ applies to the present case of a fixed number
of people $N\langle k \rangle$ living in $N$ towns: There are then $N!$ ways to associate a
given sequence of $N$ different labels with a town. However all towns with the
same $k$ are equivalent. So the number of different states are $N_{s}
=\frac{N!}{\Pi_{k}(NP(k))!}$ \ which by Stirling's approximation gives the
entropy 
\begin{equation}
S=-N\sum_{k}P(k)\ln P(k).
\end{equation}
This means that there is a one-to-one correspondence between $P(k)$ and $S$.
Thus we can change variables and instead of keeping $P(k)$ fixed we keep $S$
fixed. 

In order to find the extremum corresponding to these constraints
we need three Lagrangian multipliers $a$, $b$, and $c$ corresponding to
normalization condition for $P$, constant average number of people per town
$\langle k\rangle$, and constant entropy $S.$ Thus in these variables the
detailed balance condition determines $P(k)$ via the minimization of  the
functional \cite{G}
\begin{equation}
G[P]=\sum_{k=1}P(k)[\ln kP(k)+a+bk-c\ln P(k)] \label{G(P)} ,
\end{equation}
$\Longrightarrow \ln P(k)+\ln k+1+a+bk-c\ln P(k)-c=0$ and
hence the solution 
\begin{equation}
P(k)={\cal A}\frac{\exp(-k/\lambda)}{k^{\gamma}} \label{P(k)}
\end{equation}
with $\gamma=1/(1-c)$, $\lambda=(1-c)/b$, and ${\cal A}=\exp(1-a/(1-c))$. 
We stress that Eq.~(\ref{P(k)}) is unique and thus covers all possible
solutions for the detailed balance equation of the form Eq.~(\ref{detailed})
except solutions with order $N\langle  k\rangle$ inhabitants and other discrete
solutions which fall outside variational calculus (see below).
In order to finish the
translation into a statistical mechanical formulation we observe that each of
the possible solutions correspond to different free energies. In statistical
mechanics such solutions correspond to different temperatures $T$. The
solution for $T=0$ is the ground state and hence the absolute minimum of
$G[P(k)={\cal A}\frac{\exp(-k/\lambda)}{k^{\gamma}}]$ with respect to the
variables ${\cal A}$, $\lambda$, and $\gamma$. Minimization of $G[P]$ in
Eq.\ (\ref{G(P)}) gives the ground state solution $P(k)={\cal A}k^{-\gamma_{0}}$
with ${\cal A}$ and $\gamma_{0}$ determined from the normalization and the
constant $\langle k \rangle$ condition. Or, in other words $\gamma_{0}(\langle
k \rangle)$ is a unique function of the average town population $\langle k
\rangle$. Consequently, the Hamiltonian which gives the correct ground state is
uniquely given by
\begin{equation}
H=N\sum_{k=1}n(k)\left(  \frac{1}{\gamma_{0}}\ln n(k)+\ln k\right) ,
\label{H}
\end{equation}
where $Nn(k)$ is the number of towns with $k$ inhabitants. (Note that the
term\ $\sum_{k=1}n(k)[a+bk]$ has been dropped because it does not depend on
$n(k)$). The statistical mechanics is specified by the Boltzmann factor given
by $\exp(-H/T)$. The point is now that each one of the solutions
$P(k)=A\frac{\exp(-k/\lambda)}{k^{\gamma}}$ are recovered as an
statistical mechanical equilibrium solution; one for each temperature $T$.
This completes the statistical mechanical formulation of the town model. This
formulation is in accord with the entropy-maximum formulation by
Jaynes~\cite{jaynes} in the sense that the solution we obtain for each given
$T$ contains the maximum unbiased information you can have subject to the
imposed constraints.

\begin{figure}
\begin{center}
\includegraphics[width=0.6\columnwidth]{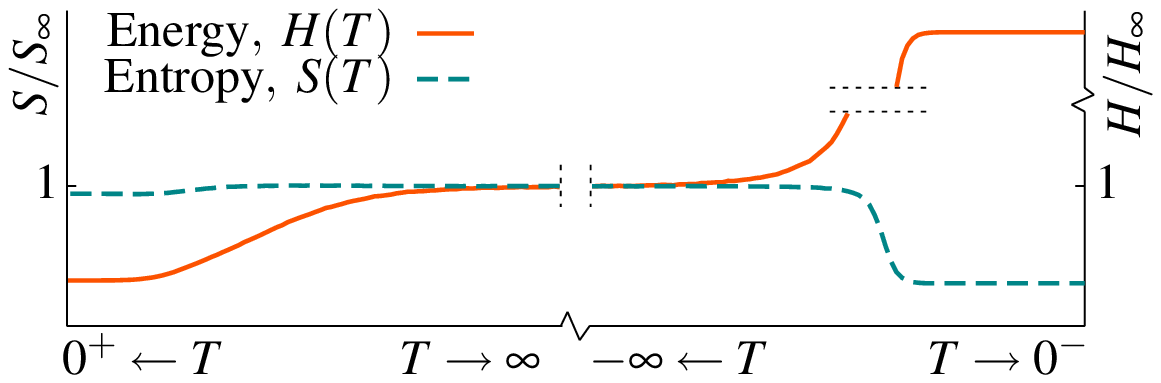}%
\caption{Energy $H(T)$ and entropy $S(T)$ as a function of $T$ for the 
town model.
For $T=0$ the entropy is finite which means that the ground state is
degenerate. As $T$ increases from $0$ to $\infty$, the entropy only increases
slightly, which means that the ground state has almost as many available states
as at $T=\infty$. On the negative-temperature side 
from $T=-\infty$ to $T=-0$ the energy increases whereas the entropy 
decreases since the number of available
states decreases. (Note the cuts and shortenings of the horizontal and
vertical axes.)}
\label{energy}
\end{center}
\end{figure}

\begin{figure}
\begin{center}
\includegraphics[width=0.6\columnwidth]{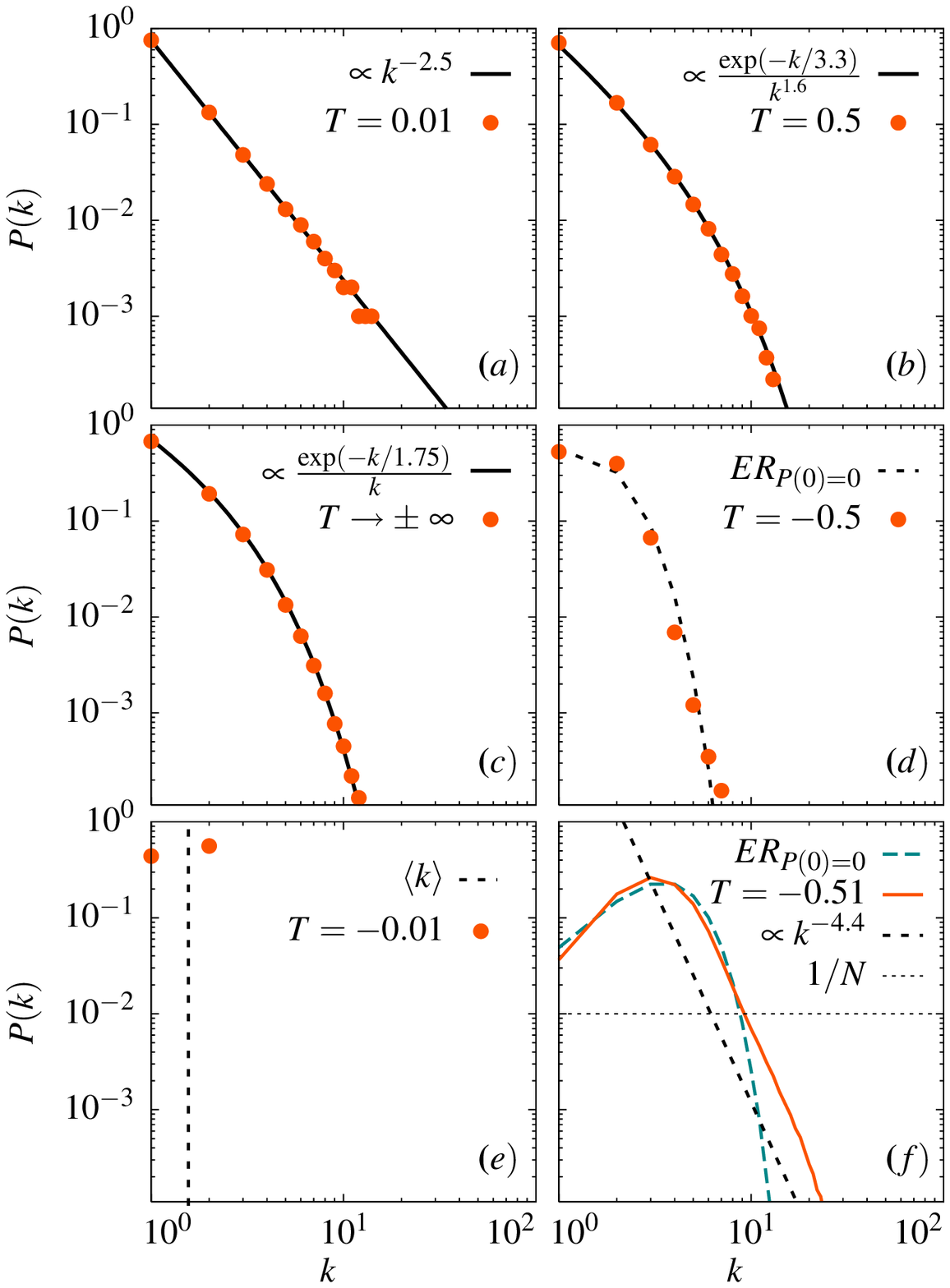}%
\caption{Distribution of town sizes $P(k)$ as a function of $T$ for $N = 1000$ and $\gamma_{0}=2.5$
which gives $\langle k\rangle=1.56$. (a) For $T\simeq0$ the distribution is scale-free. (b)-(c) As
$T$ increases from $0$ to $\infty$, the scale-free distribution gradually
changes to $P(k)\sim\exp(-k/\lambda)/k$. (d)-(e) As the
temperature is increased through $-\infty$ to $-0$, the distribution continues
to narrow and finally approaches as much as possible the single-sized town
state in (d), where the vertical dashed line denotes the solution at 
$T=-0$ of the single-sized towns of the size $\langle k \rangle=1.56$. 
The solution for $T=-0.5$ is for practical purposes very close the the 
ER-solution denoted as a dashed curve in (d), for which 
the boundary condition $P(0)=0$ is used in the ER-calculation 
for the purpose of comparison. 
(f) Comparison for the case of $N=100$ and $\langle k \rangle = 4$
between $P(k)$ at a  negative $T$ and the ER distribution: As
a larger value $\langle k \rangle=4$ is used, 
the difference is somewhat larger than in (d) for $\langle k\rangle = 1.56$. }
\label{resultP}%
\end{center}
\end{figure}

In Fig.\ \ref{energy}, we plot the energy and entropy as functions of $T$: Both
increase from $T=0$ to $T=\infty$, while from $T=-\infty$ to $T=-0$ the entropy
decreases but the energy increases. The corresponding distribution of town
sizes are illustrated in Fig.\ \ref{resultP}: It varies from the scale-free
distribution with $\lambda=0$ at $T=0$ and then narrows to the totally random
distribution with $\gamma=1$ and $\lambda>0$ at $T=\pm\infty$. On the
negative-temperature side it continues to narrow down to $T=-0$. We stress that the
Hamiltonian $H$ in Eq.~(\ref{H}) is the only Hamiltonian which contains the
complete spectrum of possible solutions consistent with the detailed balance
condition for the corresponding average distribution $P(k)$ Eq.\ (\ref{detailed}).
Also note that the Hamiltonian (\ref{H}) includes the discrete solutions 
(which cannot be obtained directly from variational calculus).

From a dynamical point of view the town model evolves in time according to a
Metropolis MC dynamics: At each time step a random person chooses another
random person and moves or does not move to the town of the latter person
according to a rate which depends on the energy difference given by the Hamiltonian. The
point we are making is that this formulation combines maximum randomness with
detailed balance \cite{palla04,park04}: Provided that the only constraint
is the detailed balance given by Eq.\ (\ref{detailed}), then our description is
unique and connects each solution $P(k)=\langle n(k) \rangle$ with the
corresponding averaged fluctuations in time
$\Delta(k)=\sqrt{ \langle n(k)^2 \rangle-\langle n(k)\rangle^2}$.
Alternatively expressed: Our Metropolis algorithm defines the unique random
rewiring process which gives rise to the average degree distribution $P(k)$
defined by Eq.~(\ref{detailed}).

Although Eq.\ (\ref{detailed}) is a quite general form of detailed balance, it
does not cover all possibilities. This is because there are alternative ways
to introduce the randomness. A second possibility is to instead choose
a random person and then a random town. This is
described by the Hamiltonian $H= N\sum_{k}n(k)\ln k!$, and the corresponding
distributions goes from single sized towns at $T=0$ to the
ER(Erd\H{o}s-R\'{e}nyi) distribution at $T=\infty$ and continues towards the
maximum entropy state $P(k)\sim\exp(-k/\lambda)$ on the negative-temperature 
side until it
collapses to a discontinuous star-type distribution \cite{bernhardsson06}:
From a statistical mechanical point of a view the system collapses from a 
low-energy and high-entropy state via a first-order transition to a 
high-energy and low-entropy state \cite{bernhardsson06}.
The third obvious possibility
is to choose two random towns, which only gives the trivial maximum entropy state $P(k)\sim\exp(-k/\lambda)$. We also
note that our randomness "random person to random person" is reminiscent of
preferential attachment since choosing a random person is equivalent to
choosing a random town with a probability $\sim kP(k)$ and then choosing one
of its inhabitants with probability $1$. We note that distributions which are
\textit{approximately} of the ER-form are obtained for a negative $T$ in
case of the random-person-to-random-person Hamiltonian as shown in 
Fig.~\ref{resultP}. The deviation comes for the largest towns as is further
illustrated in Fig.~\ref{resultP}(f). The point here is that it is in practice
difficult to distinguish between various random processes on the basis of
\textit{only} the distribution $P(k)$.%

\begin{figure}
\begin{center}
\includegraphics[width=0.6\columnwidth]{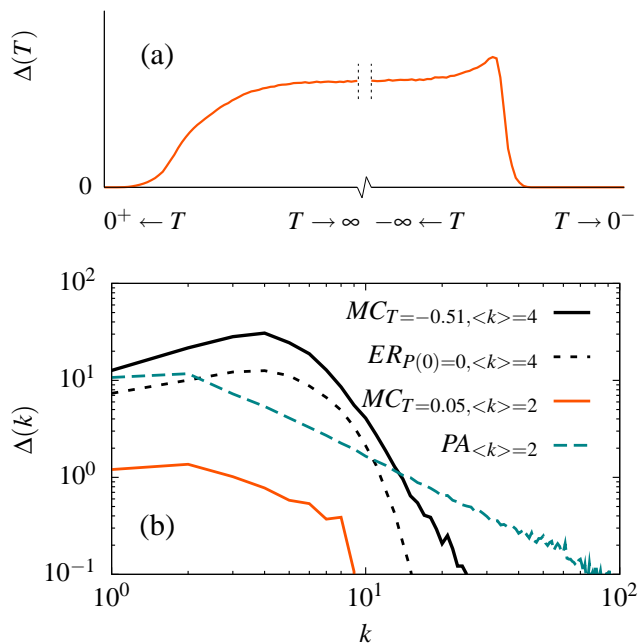}
\caption{(a) Fluctuations $\Delta(T)$ as a function of $T$. As
$T\longrightarrow+0$ the fluctuations vanish and the distribution $P(k)$
becomes scale-free. (b) The fluctuations $\Delta(k)$ as a function of the town size $k$
for a variety of distributions $P(k)$.}
\label{fluct}
\end{center}
\end{figure}

Our investigation of the town model suggests that the deviation, $\Delta(k)$,
from the average distribution during a steady state process (or more generally
any evolution process) is an essential and informative characteristics of a
network which complements the average distribution function\ $P(k)$. 
In Fig.\ \ref{fluct}(a) the total noise $\Delta=\sum_{k}\Delta(k)$ is plotted as a
function of $T$ for the town model. The most striking feature is the
vanishing of $\Delta$ as $T\rightarrow+0$, which means that the deviation from
the average $P(k)$ goes to zero. In the same limit $P(k)$ becomes scale-free 
and the number of available states is large, as discussed in connection with 
Figs.\ \ref{energy} and \ref{resultP}. As $T$ increases the noise also increases and
goes through a maximum before it drops to zero when the number of available
different states decreases to a number of order one. Figure~\ref{fluct}(b) gives
some examples of $\Delta(k)$ as a function of town size for some distributions:
"The smaller $T$ the smaller $\Delta(k)$"-connection is illustrated by the
scale-free distribution for $T=0.05$  and $\langle k \rangle=2$ in 
Fig.\ \ref{fluct}(b). This is
compared to the corresponding result for the preferential attachment, which
gives the same $P(k)$ but a larger noise. In a similar way the large noise case
for the ER-like distribution at $T=0.51$ and $\langle k \rangle=4$ is compared to the noise of the corresponding
true ER-distribution (see Fig.~\ref{resultP}(f)). As seen 
from Fig. \ref{fluct}(b) both have large deviations from the average distribution, although the ER-deviation is somewhat smaller.

We have shown that the detailed balance under rather general conditions leads
to a unique equilibrium Hamiltonian which connects the distribution function
$P(k)$ to a fluctuation distribution $\Delta(k)$. In principle, this means that
given a distribution $P(k)$ the corresponding fluctuation distribution will
tell whether or not the dynamics of the system is consistent with detailed
balance under the specified conditions. Thus the fluctuation distribution
$\Delta(k)$ is a characteristics which complements the average distribution
$P(k)$. Preliminary tests on some real networks suggest that this might be a
useful characteristics in practice~\cite{bernhardsson06}. In addition, we have
shown that the ground state of the Hamiltonian for the town model is the highly
degenerate scale-free distribution. This state is characterized by many
possible states combined with very small fluctuations. We speculate that, in as
far as large changes and fluctuations might be fatal and many different
possibilities beneficial for the evolution of a system, the scale-free
distribution might sometimes be an evolutionary winner.

B.J.K. was supported by grant No. R01-2005-000-10199-0
from the Basic Research Program of the Korea Science
and Engineering Foundation,
P.M. and S.B. from Swedish VR contract 50412501.


\begin{thebibliography}{99}
\bibitem{footnote} Both a growing network and preferential attachment of added
links are required for the preferential attachment scheme to produce a
scale-free distribution. See, e.g.,~\cite{albert02}.
\bibitem{footnote2} Strictly speaking, it means that at each time step only a
small fraction of the links can be simultaneously moved. This fraction has to
be small enough for the chance of randomly picking two link-ends on the same
node to be insignificant.
\bibitem{G}
We are free to choose the sign of $G[P]$ without affecting the content of the formulation.
\end{thebibliography}
\end{document}